# Slot-waveguide cavities for optical quantum information applications

Mark P. Hiscocks[1], Chun-Hsu Su[2], Brant C. Gibson[2], Andrew D. Greentree[2], Lloyd C. L. Hollenberg[2], and François Ladouceur[1]

[1]*School of EE&T, UNSW, Sydney, NSW 2052, Australia*
[2]*Quantum Communications Victoria, School of Physics, University of Melbourne, Melbourne, VIC 3010, Australia*
*Corresponding author: m.hiscocks@student.unsw.edu.au*

**Abstract:** To take existing quantum optical experiments and devices into more practical regimes requires the construction of robust, solid-state implementations. In particular, to observe the strong-coupling regime of atom-photon interactions requires very small cavities and large quality factors. Here we show that the slot-waveguide geometry recently introduced for photonic applications is also promising for quantum optical applications in the visible regime. We study diamond- and GaP-based slot-waveguide cavities (SWCs) compatible with diamond colour centres e.g. nitrogen-vacancy (NV) defect. We show that one can achieve increased single-photon Rabi frequencies of order $O(10^{11})$ rad s$^{-1}$ in ultra-small cavity modal volumes, nearly 2 orders of magnitude smaller than previously studied diamond-based photonic crystal cavities.

**References and links**

1. D. F. Walls and G. J. Milburn, *Quantum Optics* (Springer-Verlag, New York, 1995).
2. M. J. Hartmann, F. G. S. L. Brandão, and M. B. Plenio, "Strongly interacting polaritons in coupled arrays of cavities," Nature Phys. **2**, 849–855 (2006).
3. A. D. Greentree, C. Tahan, J. H. Cole, and L. C. L. Hollenberg, "Quantum phase transitions of light," Nature Phys. **2**, 856–861 (2006).
4. D. G. Angelakis, M. F. Santos, and S. Bose, "Photon-blockade-induced Mott transitions and XY spin models in coupled cavity arrays," Phys. Rev. A **76**, 031805(R) (2007).
5. S. M. Spillane, T. J. Kippenberg, and K. J. Vahala, "Ultralow-threshold Raman laser using a spherical dielectric microcavity," Nature **415**, 621–623 (2002).
6. C. K. Law and H. J. Kimble, "Deterministic generation of a bit-stream of single-photon pulses," J. Mod. Opt. **44**, 2067–2074 (1997).
7. X. Maître, E. Hagley, G. Nogues, C. Wunderlich, P. Goy, M. Brune, J. M. Raimond, and S. Haroche, "Quantum memory with a single photon in a cavity," Phys. Rev. Lett. **79**, 769–772 (1997).
8. T. Pellizzari, S. A. Gardiner, J. I. Cirac, and P. Zoller, "Decoherence, continuous observation, and quantum computing: A cavity QED model," Phys. Rev. Lett. **75**, 3788–791 (1995);
9. L.-M. Duan, M. D. Lukin, J. I. Cirac, and P. Zoller, "Long-distance quantum communication with atomic ensembles and linear optics," Nature **414**, 413–418 (2001).
10. K. J. Vahala, "Optical microcavities," Nature **424**, 839–846 (2003).
11. S. Noda, M. Fujita, and T. Asano, "Spontaneous-emission control by photonic crystals and nanocavities," Nature Phot. **1**, 449–458 (2007).
12. A. D. Greentree, B. A. Fairchild, F. M. Hossain, and S. Prawer, "Diamond integrated quantum photonics," Mater. Today **11**, 22–31 (2008).
13. C. Kurtsiefer, S. Mayer, P. Zarda, and H. Weinfurter, "Stable solid-state source of single-photons," Phys. Rev. Lett. **85**, 290–293 (2000).
14. A. Beveratos, R. Brouri, T. Gacoin, A. Villing, J.-P. Poizat, and P. Grangier, "Single photon quantum cryptography," Phys. Rev. Lett. **89**, 187901 (2002).
15. F. Jelezko, T. Gaebel, I. Popa, M. Domham, A. Gruber, and J. Wrachtrup, "Observation of coherent oscillation of a single nuclear spin and realization of a two-qubit conditional quantum gate," Phys. Rev. Lett. **93**, 130501 (2004).

16. T. Gaebel, M. Domhan, I. Popa, C. Wittmann, P. Neumann, F. Jelezko, J. R. Rabeau, N. Stavrias, A. D. Greentree, S. Prawer, J. Meijer, J. Twamley, P. R. Hemmer, and J. Wrachtrup, "Room-temperature coherent coupling of single spins in diamond," Nature Phys. **2**, 408–413 (2006).
17. C. L. Degen, "Scanning magnetic field microscope with a diamond single spin sensor," Appl. Phys. Lett. **92**, 243111 (2008).
18. J. R. Maze, P. L. Stanwix, J. S. Hodges, S. Hong, J. M. Taylor, P. Cappellaro, L. Jiang, M. V. Gurudev Dutt, E. Togan, A. S. Zibrov, A. Yacoby, R. L. Walsworth, and M. D. Lukin, "Nanoscale magnetic sensing with an individual electronic spin in diamond," Nature **455**, 644–647 (2008).
19. G. Balasubramanian, I. Y. Chan, R. Kolesov, M. Al-Hmoud, J. Tisler, C. Shin, C. Kim, A. Wojcik, P. R. Hemmer, A. Krüger, T. Hanke, A. Leitenstorfer, R. Bratschitsch, F. Jelezko, and J. Wrachtrup, "Nanoscale imaging magnetometry with diamond spins under ambient conditions," Nature **455**, 648–651 (2008).
20. J. H. Cole and L. C. L. Hollenberg, "Scanning quantum decoherence microscopy," http://arxiv.org/abs/0811.1913 (2008).
21. C.-H. Su, A. D. Greentree, and L. C. L. Hollenberg, "Towards a picosecond transform-limited nitrogen-vacancy based single-photon source," Opt. Express **16**, 6240 (2008).
22. T. P Spiller, K. Nemoto, S. L. Braunstein, W. J. Munro, P. van Loock, and G. J. Milburn, "Quantum computation by communication," New J. Phys. **8**, 30 (2006).
23. C.-H. Su, A. D. Greentree, W. J. Munro, K. Nemoto, and L. C. L. Hollenberg, "High-speed quantum gates with cavity quantum electrodynamics," Phys. Rev. A **78**, 062336 (2008).
24. A. D. Greentree, J. Salzman, S. Prawer, and L. C. L. Hollenberg, Quantum gate for $Q$ switching in monolithic photonic-band-gap cavities containing two-level atoms," Phys. Rev. A **73**, 013818 (2006).
25. A. M. Stephens, Z. W. E. Evans, S. J. Devitt, A. D. Greentree, A. G. Fowler, W. J. Munro, J. L. O'Brien, K. Nemoto, and L. C. L. Hollenberg, "Deterministic optical quantum computer using photonic modules," Phys. Rev. A **78**, 032318 (2008).
26. S. J. Devitt, A. G. Fowler, A. M. Stephens, A. D. Greentree, L. C. L. Hollenberg, W. J. Munro, and K. Nemoto, "Topological cluster state computation with photons," http://arxiv.org/abs/0808.1782 (2008).
27. G. Davies and M. F. Hamer, "Optical studies of the 1.945eV vibronic band in diamond," Proc. R. Soc. Lond. A: Math. and Phys. Sci. **348,** 285–298 (1976).
28. C. F. Wang, R. Hanson, D. D. Awschalom, E. L. Hu, T. Feygelson, J. Yang, and J. E. Butler, "Fabrication and characterization of two-dimensional photonic crystal microcavities in nanocrystalline diamond," Appl. Phys. Lett. **91**, 201112 (2007).
29. S. Tomljenovic-Hanic, M. J. Steel, C. Martijn de Sterke, and J. Salzman, "Diamond based photonic crystal microcavities," Opt. Express **14**, 3556–3562 (2006).
30. I. Bayn and J. Salzman, "High-Q photonic crystal nanocavities on diamond for quantum electrodynamics," Eur. Phys. J. Appl. Phys. **37**, 19–24 (2007).
31. C. Kreuzer, J. Riedrich-Möller, E. Neu, and C. Becher, "Design of photonic crystal microcavities in diamond films," Opt. Express **16**, 1632–1644 (2008).
32. M. W. McCutcheon and M. Lončar, "Design of a silicon nitride photonic crystal nanocavity with a Quality factor of one million for coupling to a diamond nanocrystal," Opt. Express **16**, 19137–19145 (2008).
33. I. Bayn, and J. Salzman, "Ultra high-Q photonic crystal nanocavity design: "The effect of a low-ε slab material," Opt. Express **16**, 4972–4980 (2008).
34. S. Tomljenovic-Hanic, A. D. Greentree, C. Martijn de Sterke, and S. Prawer, "Design of flexible ultrahigh-Q microcavities in diamond-based photonic crystal slabs," Opt. Express **17**, 6465–6475 (2009).
35. E. Wu, J. R. Rabeau, G. Roger, F. Treussart, H. Zeng, P. Grangier, S. Prawer, and J.-F. Roch, "Room temperature triggered single-photon source in the near infrared," New J. Phys. **9**, 434 (2007).
36. C. Wang, C. Kurtsiefer, H. Weinfurter, and B. Burchard, "Single photon emission from SiV centres in diamond produced by ion implantation," J. Phys. B: At. Mol. Opt. Phys. **39**, 37–41 (2006).
37. I. Aharonovich, C. Zhou, A. Stacey, J. Orwa, D. Simpson, A. D. Greentree, F. Treussart, J.-F. Roch, and S. Prawer, "A new, enhanced diamond single photon emitter in the near infra-red," http://arxiv.org/abs/0902.3051 (2009).
38. V. R. Almeida, Q. Xu, C. A. Barrios, and M. Lipson, "Guiding and confining light in void nanostructure," Opt. Lett. **29**, 1209 (2004).
39. Q. Xu, V. R. Almeida, R. R. Panepucci, and M. Lipson, "Experimental demonstration of guiding and confining light in nanometer-size low-refractive-index material," Opt. Lett. **29**, 1626–1628 (2004).
40. J. T. Robinson, C. Manolatou, L. Chen, and M. Lipson, "Ultrasmall mode volumes in dielectric optical microcavities," Phys. Rev. Lett. **95**, 143901 (2005).
41. R. Sun, P. Dong, N. Feng, C. Hong, J. Michel, M. Lipson, and L. Kimerling, "Horizontal single and multiple slot waveguides: optical transmission at λ = 1550 nm," Opt. Express **15**, 17967–17972 (2007).
42. A. Gondarenko and M. Lipson, "Low modal volume dipole-like dielectric slab resonator," Opt. Express **16**, 17689 (2008).
43. M. W. Pruessner, T. H. Stievater, and W. S. Rabinovich, "Integrated waveguide Fabry–Perot microcavities with silicon/air Bragg mirrors," Opt. Lett. **32**, 533-535 (2007).
44. P. Velha, E. Picard, T. Charvolin, E. Hadji1, J. C. Rodier, P. Lalanne and D. Peyrade, "Ultra-High Q/V Fabry-Perot microcavity on SOI substrate," Opt. Express **15**, 16090–16096 (2007).

45. M. O. Scully and M. S. Zubairy, *Quantum Optics* (Cambridge, 1997).
46. N. B. Manson, J. P. Harrison, and M. J. Sellars, "Nitrogen-vacancy center in diamond: Model of the electronic structure and associated dynamics," Phys. Rev. B **74**, 104303 (2006).
47. D. E. Aspnes and A. A. Studna, "Dielectric functions and optical parameters of Si, Ge, GaP, GaAs, GaSb, InP, InAs, and InSb from 1.5 to 6.0 eV," Phys. Rev. B **27**, 985–1009 (1983).
48. FIMMWAVE, Photon Design, http://www.photond.com.
49. G. Cui and M. Raymer, "Quantum efficiency of single-photon sources in the cavity-QED strong-coupling regime," Opt. Express **13**, 9660–9665 (2005).
50. K.-M. C. Fu, C. Santori, P. E. Barclay, I. Aharonovich, S. Prawer, N. Meyer, A. M. Holm, and R. G. Beausoleil, "Coupling of nitrogen-vacancy centers in diamond to a GaP waveguide," Appl. Phys. Lett. **93**, 234107 (2008).
51. J. Butler and A. V. Sumant, "The CVD of nanodiamond materials," Chem. Vapor Depos. **14**, 145–160 (2008).
52. P. Olivero, S. Rubanov, P. Reichart, B. C. Gibson, S. T. Huntington, J. Rabeau, A. D. Greentree, J. Salzman, D. Moore, D. N. Jamieson, S. Prawer, "Ion-beam-assisted lift-off technique for three-dimensional micromachining of freestanding single-crystal diamond," Adv. Mater. **17**, 2427-2430 (2005).
53. B. A. Fairchild, P. Olivero, S. Rubanov, A. D. Greentree, F. Waldermann, R. A. Taylor, I. Walmsley, J. M. Smith, S. Huntington, B. C. Gibson, D. N. Jamieson and S. Prawer, "Fabrication of ultrathin single-crystal diamond membranes," Adv. Mater. **20**, 4793-4798 (2008).
54. M. P. Hiscocks, K. Ganesan, B. C. Gibson, S. T. Huntington, F. Ladouceur and S. Prawer, "Diamond waveguides fabricated by reactive ion etching," Opt. Express **16**, 19512–19519 (2008).

## 1. Introduction

Cavities and optical resonators are invaluable test beds for the study of quantum optics [1] and condensed matter analog systems [2–4]. They are the building blocks for prospective technologies including nonlinear optics [5], single-photon sources [6], quantum memories [7] and quantum information processing [8–9]. For these applications, it is critical to realize cavities with both a high quality ($Q$) factor and small mode volume $V$ since the ratio $Q/V$ determines the strength and coherence time of various cavity interactions, and large-scale integrability of the device.

Numerous cavity designs such as micropillars, microspheres, microdisks, and defects in photonic crystals have been extensively investigated and experimentally demonstrated [10]. In particular, significant attention has turned towards these photonic-bandgap (PBG) defect cavities owing to their extremely high $Q/V$ ratio where experimentally, $Q = 10^6$ for a silicon-based structure operating at near-infrared (IR) wavelengths has been achieved [11]. However there are few high-$Q/V$ schemes that are compatible with a wide array of diamond-based, quantum electrodynamics (QED) proposals [12] in the visible regime. Optically active diamond defects such as nitrogen-vacancy (NV) centres have shown suitability for single-photon emission [13], quantum communication [14], quantum information processing [15,16], and nano-scale magnetometry [17–20]. When matched by high quality cavities, opportunities arise for cavity-QED with NV centres, allowing enhancement of their optical transitions [21] and remote qubit-qubit interaction [22,23]. Cavity-assisted photonic networks [24] based on NV centres for large-scale quantum computing have also been discussed [25,26]. These possibilities motivate the recent demonstration of PBG cavity modes that resonate near the NV centre's zero-phonon line (ZPL, $\lambda = 637$ nm) [27] with $Q = 585$ [28] and proposals for larger-$Q$ PBG cavities in diamond [29–34]. Other diamond defects such as the nickel-related NE8 ($\lambda = 794$ nm) [35] and silicon-vacancy (SiV, 737 nm) centres [36], and some newer, uncharacterized centres [37], which show sharp emission lines and large transition dipole moments, are also promising candidates for photon-emission improvement with cavities.

Within this context, we investigate a newly developed class of waveguides – slot waveguides [38] with the purpose of utilizing slot-waveguide cavities (SWCs) for applications with diamond centres in the visible regime. As depicted in Fig. 1(a), the light is strongly guided in the narrow region (i.e. slot) of lower refractive index $n_S$ between two regions (rods) of higher refractive index $n_R$. The large discontinuities at high index-contrast interfaces enable this light confinement. To date, existing proposals and demonstrations have been limited to near-IR operations [39–42]. Here we show that slot-waveguide approaches can be used to enhance diamond colour centres for quantum applications. By combining with mirrors, PBG or distributed Bragg reflectors (DBRs) in a Fabry-Perot arrangement as shown in Fig. 1(b), diamond- and GaP-based slot structures define suitable cavities with modal volumes up to nearly 2 orders of magnitude smaller than previously studied wavelength-sized photonic crystal cavities. This predicts effective cavity-NV single-photon Rabi frequency of $202 \times 10^9$ rad s$^{-1}$ and $135 \times 10^9$ rad s$^{-1}$ in GaP-air and all-diamond slot structures, respectively. Both DBRs and PBG reflectors capable of achieving cavity-Q factor more than $10^4$ have been demonstrated [43,44]. These results represent a significant step in improving the prospects for solid-state cavity-QED based quantum logic in diamond.

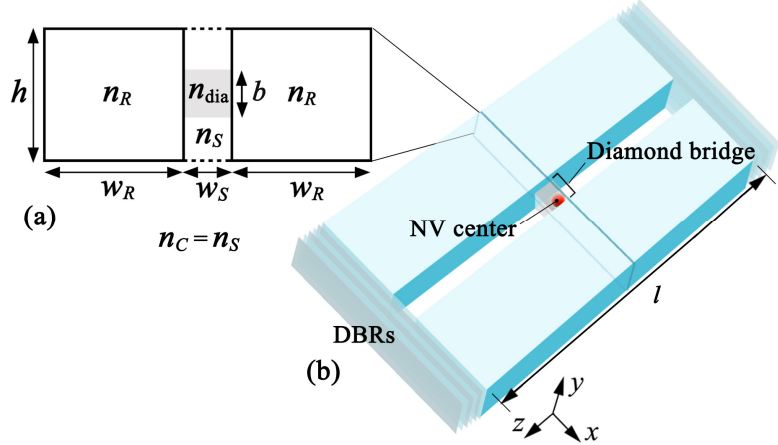

Fig. 1. Schematic of a slot-waveguide cavity (SWC). (a) Cross-section through the slot-waveguide structure in the *xy*-plane going through the diamond bridge, which is centrally located as indicated in the diagram. Note the rectangular slots of high-index material defining the cavity-structure and the smaller diamond region. (b) Slot waveguide combined with distributed Bragg reflectors defines a resonant structure with length $l = \lambda/2$. A diamond optical centre, e.g. NV defect, housed in a nanocrystal or diamond bridge can be coupled to the cavity mode.

## 2. Cavity designs

To achieve strong, cavity-assisted atom-photon interactions, we require that the electric field *per photon* be maximal in the vicinity of the dipole. It is generally accepted that the solution of the classical Maxwell's equations can be reinterpreted as a precise description of a one-photon state. Thus, we can classically calculate the mode field distribution $E(\mathbf{r})$ for a given cavity mode and rescale appropriately. To do so, we define – as an intermediate quantity – the (classically calculated) energy density $\xi(\mathbf{r}) = n(\mathbf{r})^2|E(\mathbf{r})|^2/[n(\boldsymbol{\rho})|E(\boldsymbol{\rho})|]^2$, where $\boldsymbol{\rho}$ is the point where $n(\mathbf{r})^2|E(\mathbf{r})|^2$ is maximum. The per-photon electric field can then be expressed by [45]:

$$|\mathcal{E}(\mathbf{r})| = \sqrt{\frac{\hbar\omega_i \xi(\mathbf{r})}{2\varepsilon_0 n(\mathbf{r})^2 V}}, \quad (1)$$

where we have also introduced the mode volume $V$ given by:

$$V = \frac{\int n(\mathbf{r})^2 |E(\mathbf{r})|^2 \, d^3\mathbf{r}}{n(\boldsymbol{\rho})^2 |E(\boldsymbol{\rho})|^2}. \quad (2)$$

As an example, we consider the simple case of a point defect cavity in a diamond-based PBG with a mode volume $(\lambda/n)^3$, which is of the order typically obtained in theoretical diamond designs [29]. The theoretical limit for the fundamental mode of a point defect cavity is of order $(\lambda/2n)^3$ but these fundamental modes are normally not used because they are lossy [11]. For a cavity resonant wavelength of 637 nm, the maximum single-photon field amplitude in the cavity is 0.4 MV/m. By way of contrast, we will use this field and the standard mode volume as our benchmark, against which we compare our SWC designs.

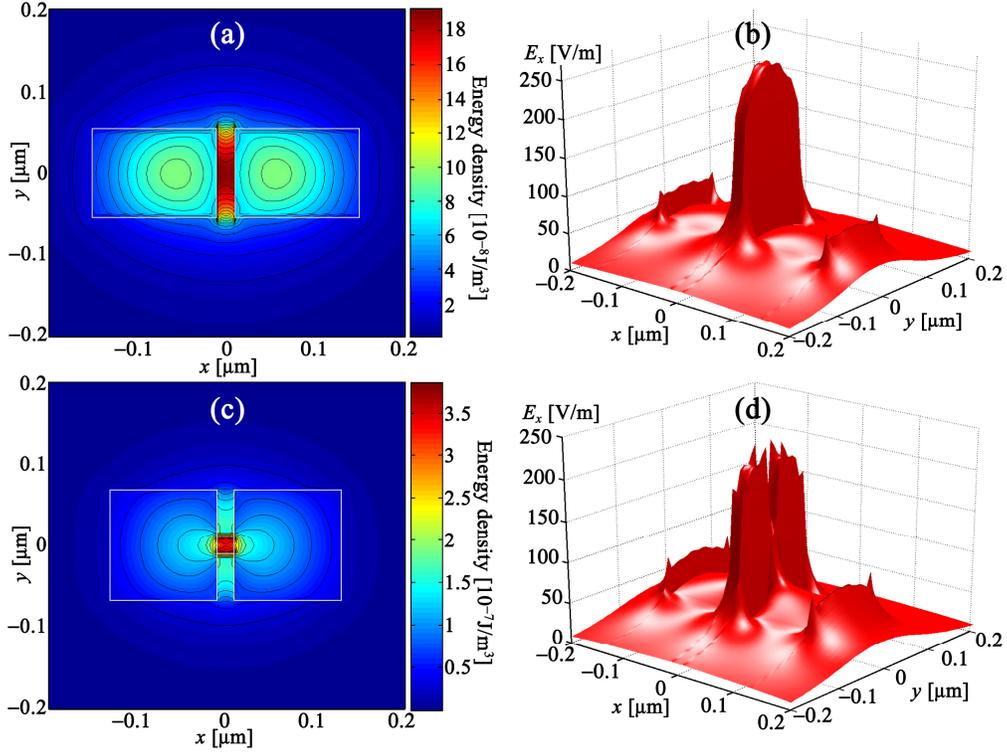

Fig. 2. Fundamental quasi-TE mode of diamond-air slot waveguides. (a,c) Energy density, (b,d) $E_x$ distributions for fundamental mode for a $\{w_S, w_R, h\} = \{20,140,110\}$ nm empty slot and a $\{20,120,130\}$ nm slot with a 20 nm high, cavity-long bridge respectively. The $x$-component of respective per-photon electric field amplitudes, $\mathcal{E}_x$, at slot or bridge centre are given in Table 1. Note the reduction in electric field in the vicinity of the bridge in (d), however this is still greater than for the $\lambda^3$-cavity case.

An electric dipole in the cavity will couple to the electric field, resulting in observable Rabi oscillations providing the dipole-cavity coupling is greater than the competing loss processes (e.g. absorption and photon outcoupling). The single-photon Rabi frequency is given by $g(\mathbf{r}) = \boldsymbol{\mu} \cdot \boldsymbol{\mathcal{E}}(\mathbf{r})/\hbar$ in units of angular frequency, where $\boldsymbol{\mu}$ is the transition dipole moment associated with the atomic transition of frequency $\omega$. It is therefore a measure of the coupling strength of the matter-photon interaction on the quantum level. The magnitude of the transition dipole moment can be determined from the spontaneous emission rate $\gamma$ using the relation $|\boldsymbol{\mu}|^2 = 3\pi\hbar\varepsilon_0 c^3 \gamma/(n\omega^3)$. For instance, the NV centre has an excited-level lifetime of around 11.6 ns [46] and emission via the purely electronic ZPL decay channel contributes 3% of the total photoluminescence output [27,46]. The relatively low branching ratio to the ZPL transition lowers the effective transition dipole moment of the ZPL to $3.1 \times 10^{-30}$ Cm (compared to $1.8 \times 10^{-29}$ Cm for the total transition dipole moment), leading to a single-photon Rabi frequency of $12 \times 10^9$ rad s$^{-1}$ for a $\lambda^3$ cavity.

We consider four designs, (*1*) diamond-air structure: using diamond ($n_R = n_{\text{dia}} = 2.4$) as the high index rods and air ($n_S = n_{\text{air}} = 1$) as the low index slot, (*2*) GaP-air: GaP rods ($n_R = n_{\text{GaP}} = 3.3$) with an air slot, (*3*) diamond-silica: diamond rods and silica slot ($n_S = n_{\text{silica}} = 1.45$), and (*4*) GaP-diamond: GaP rods and diamond-slot. The claddings of the structures are taken to be matched the slot material. These media are chosen for their transparency in the visible. Diamond possesses the widest optical transparency window, GaP is transparent over the 554–828 nm range [47], and cavity loss via absorption in silica is only

$6 \times 10^{-5}$ dB/cm, hence they are suitable for our purposes. They also set up required index discontinuities that enable cavity confinements. The GaP-air slot should exhibit the greatest confinement due to a large index contrast $(n_R - n_S)/n_R = 0.70$, followed by (in order) diamond-air (contrast is 0.59), diamond-silica (0.40) and GaP-diamond (0.28) slots. We solve for the fundamental quasi-TE mode of these structures at $\lambda = 637$ nm numerically using the FIMMWAVE FMM real solver [48]. Since these true modes are theoretically lossless and materials are near non-attenuating, cavity loss is predominantly due to scattering at the reflective boundaries. The characterization of cavity quality factor $Q$ of a SWC is beyond the scope of this work but we comment on the state-of-the-art engineering in Section 5.

Finally, some of these calculated cavity modes are reported in Fig. 2. In Figs. 2(a,b), the optical intensity and the $E_x$-field distributions of the TE-mode are calculated for a diamond-air structure with dimensions $\{w_S, w_R, h\} = \{20,140,110\}$ nm, illustrating a strong E-field confinement within the empty slot region. In the first three designs, the NV centre must ultimately be housed in a nanocrystal or diamond bridge. If the nanocrystal or bridge is small compared to rod height $h$ and cavity length $l$, it will only induce a small perturbation to the local field of the standard mode. These empty-slot solutions can be taken as representative mode solutions for a SWC augmented with a small nanocrystal or bridge. In Figs. 2(c,d), we consider the use of a 20 nm high bridge that extends over the cavity length inside a $\{20,120,130\}$ nm diamond-air slot. Despite the reduction in field intensity in the vicinity of the bridge, the local field and energy density remain high within the higher-$n$ diamond bridge and, more importantly as we will see in next section, are still greater than the comparable $\lambda^3$-cavities.

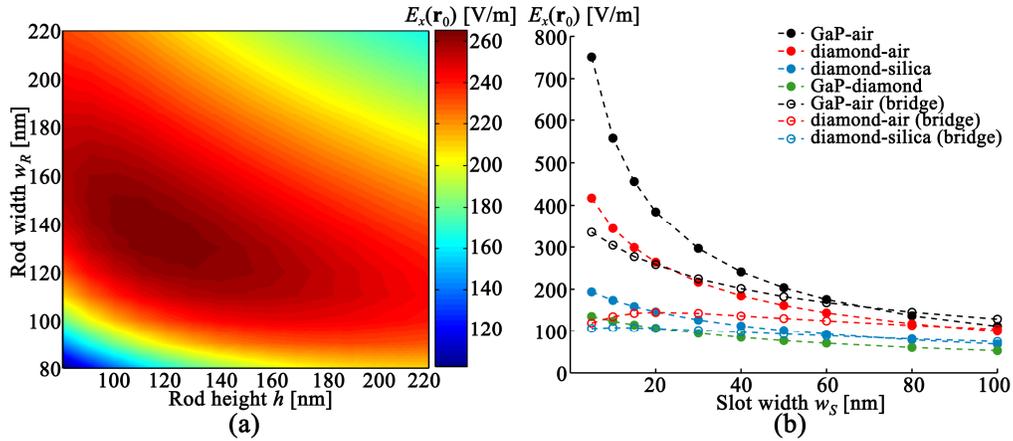

Fig. 3. (a) Field amplitude $E_x(\mathbf{r}_0)$ for varying rod dimensions of a diamond-air structure with slot width $w_S = 20$ nm. (b) $E_x(\mathbf{r}_0)$ for different $w_S$ and various slot waveguide designs using optimal rod designs given in Table 1. Hollow and solid circles denote data points for designs with and without a 20 nm diamond bridge, respectively. Dashed lines are guides for eyes.

## 3. Design optimization

We first investigate the influence of rod width $w_R$ and height $h$ on the dominant $E_x(\mathbf{r})$ component of the TE-mode. In Fig. 3(a), where this is calculated for a 20 nm slot diamond-air arrangement, there is an optimal rod size of roughly $140 \times 110$ nm ($w_R \times h$) that maximizes the field strength $E_x(\mathbf{r}_0)$, where $\mathbf{r}_0$ is the centre of the cavity and the assumed location of the dipole. This optimal design is dependent on the slot width, although only weakly. We have used the optimal 20 nm design as a close match over the range we are investigating. The simulations are performed for 1 fW input power using a domain size of $2.36 \times 2$ μm. Significant variations in rod dimensions must occur before the local field is halved. The same optimization routine was performed for the other designs with a 20 nm slot. The optimum rod dimensions, and corresponding $E_x(\mathbf{r}_0)$ and $x$-component of the per-photon field $\mathcal{E}_x(\mathbf{r}_0)$, are summarized in Table 1. Notably, the field amplitudes are an order of magnitude greater than

0.4 MV/m in the $\lambda^3$-cavity. Further mode improvements are expected by reducing the slot width, as shown in Figs. 3(b) and 4. When the bridge is absent, decreasing slot width (< 30 nm), of particularly higher-contrast structures (i.e. GaP-air and diamond-air), results in a greater increase in $E_x(\mathbf{r}_0)$ and improvement in optical confinement. This trend is not shared by the structures with a bridge – the variations in $E_x(\mathbf{r}_0)$ are relatively small with decreasing slot width. In other words, the bridge has the advantage of guiding the light field strongly between the rods, as shown in Fig. 2(c), even when the slot is wide. Although the 5 nm slot GaP-air arrangement represents the better design, the practicality of realizing this scheme in comparison with other arrangements warrants further discussion in Section 5.

Table 1. Optimal designs for maximizing $E_x(\mathbf{r}_0)$ in various slot-waveguides with 20 nm slot width.

| Rod material | Slot/Cladding material | Bridge height $b$ [nm] | Optimum rod size $(w_R \times h)$ [nm] | $E_x(\mathbf{r_0})$ [V/m] | $\mathcal{E}_x(\mathbf{r_0})$ [MV/m] |
|---|---|---|---|---|---|
| Diamond | Air | N/A | 140 × 110 | 263.9 | 3.0 |
| GaP | Air | N/A | 110 × 80* | 384.3 | 3.7 |
| GaP | Diamond | N/A | 90 × 120 | 107.0 | 1.0 |
| Diamond | Silica | N/A | 130 × 130 | 146.9 | 2.8 |
| Diamond | Air | 20 | 120 × 130 | 145.0 | 1.6 |
| GaP | Air | 20 | 90 × 90 | 258.5 | 2.3 |
| Diamond | Silica | 20 | 120 × 140 | 101.7 | 0.9 |

* All other listed slot-waveguide designs are single moded for their optimum design. For the GaP-air arrangement, rod size of 110 × 70 nm gives the highest $E_x(\mathbf{r}_0)$ whilst remaining single-moded.

## 4. Cavity-dipole interaction

Next, we consider the use of an NV-coupled SWC for cavity-QED applications. By assuming a SWC of length $l = \lambda/2$ and a field distribution of the form $E_x(\mathbf{r})\sin(2\pi z/\lambda)$, we plot the normalized volume $V(n_{\text{dia}}/\lambda)^3$ in Fig. 4(a) and the corresponding per-photon field amplitudes $\mathcal{E}_x$ and Rabi frequency with the NV's ZPL transition in Fig. 4(b) for different slot widths. These results generally reflect the trends observed in our $E_x$ results, e.g. the mode volume decreases with smaller slots and larger local field amplitude and stronger matter-photon interaction as measured by Rabi frequency.

The GaP-air SWC can achieve $V = 0.02(\lambda/n_{\text{dia}})^3$ with a 5 nm slot and the ratio $\xi(\mathbf{r}_0) = 0.96$ at the centre. Hence we expect a single-photon Rabi frequency $g(\mathbf{r}_0)$ of $202 \times 10^9$ rad s$^{-1}$ for 110 × 70 nm GaP rods, and $110 \times 10^9$ rad s$^{-1}$ when $V = 0.05(\lambda/n_{\text{dia}})^3$ for a 20 nm wide slot. We note that similar reductions in mode volumes are reported in Refs. [40,42]. If the cavity-long bridge and square 90 × 90 nm rod are employed, the couplings are $85 \times 10^9$ and $70 \times 10^9$ rad s$^{-1}$, respectively. Similarly, an all-diamond cavity also enables $g(\mathbf{r}_0) = 90 \times 10^9$ rad s$^{-1}$ for 20 nm slot and $47 \times 10^9$ rad s$^{-1}$ if the bridge is present. In comparison, these matter-photon couplings are now 4 to 20-fold stronger than that achievable in a $\lambda^3$-cavity. Thus the enhanced strength of the coherent dynamics to the dissipative processes, characterized by atomic cooperativity parameter $C \equiv g(\mathbf{r}_0)^2/(\Gamma\kappa)$ has values up to $\sim 5 \times 10^{14}/\kappa$ where $\kappa = \omega/(2Q)$ is the cavity decay rate and $1/\Gamma$ the atomic lifetime. For the NV where $\omega = 2.95 \times 10^{15}$ rad s$^{-1}$, the regime of $C \gg 1$ can be achieved with modest $Q$ values. To use the system as a high-performance single-photon source in the weak coupling regime of cavity-QED, the cavity boundaries should define a modest $Q$ of $10^3$–$10^4$ or cavity decay rate of $10^2$–$10^3 \times 10^9$ rad s$^{-1}$ for resonant Purcell enhancement of factor $3(\lambda/n_{\text{dia}})^3 \times Q/(4\pi^2 V) \leq O(10^4)$ – a near unit quantum efficiency $[g(\mathbf{r}_0)^2/(g(\mathbf{r}_0)^2 + \kappa\Gamma)][\kappa/(\kappa + \Gamma)] \leq 0.999$ for single photon output in picosecond time scale [21,49]. Furthermore, as the requirement on $Q$ is relaxed, the strong coupling regime $\kappa < 2g(\mathbf{r}_0)$ becomes more accessible in the slot designs than the $\lambda^3$-cavity counterpart.

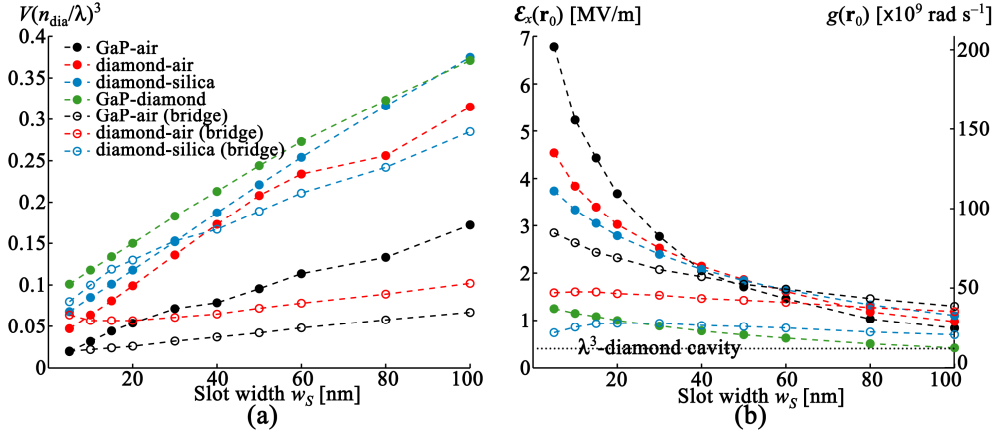

Fig. 4. (a) Mode volumes of fundamental quasi-TE mode normalized by the typical volume of a diamond-based $\lambda^3$-cavity, $(\lambda/n_{dia})^3$ and (b) Normalized per-photon electric field amplitude $\mathcal{E}_x$ and Rabi frequency $g(\mathbf{r_0})$ with the ZPL transition of the NV, as a function of slot width $w_S$ for various slot-waveguide designs. The rod specifications and notations follow Fig. 3(b). The dotted line denotes the corresponding field and coupling strength for the $\lambda^3$-cavity.

### 5. Fabrication considerations

Of the designs we have proposed, the most straightforward to fabricate would be the GaP-air or diamond-air structures, mainly through the simplicity afforded by a monolithic structure. Vertical confinement could be achieved for GaP structures by transferring a GaP layer onto a substrate of silica or diamond [50]. For diamond there is the possibility of depositing CVD diamond onto silica [51] or more desirably using ion implantation to create an airgap or membrane in single crystal diamond [52,53]. The waveguides could then be fabricated in these layers using e-beam lithography (EBL) to pattern the structures [38] and then using the appropriate masking/dry etching techniques for the material [54]. Inclusion of the NV centre in the structure can be achieved using either a diamond nanocrystal or the inclusion of a diamond bridge. Both solutions present fabrication difficulties although the nanocrystal seems the easier of the two. A nanocrystal will require careful placement in the air-slot and ideally have some method of positioning it at the centre by height. The diamond bridge adds significant complexity to the fabrication, but it may be possible to apply the ion implantation methods mentioned above to create the structure entirely in single crystal diamond (i.e. diamond rods and bridge) or perhaps by more standard deposition methods in a mixed material scenario.

Although the slot designs presented earlier in this paper have a vertical slot, fabricating the slot horizontally [41] has a number of advantages. These include reducing interface roughness and facilitating narrower slots due to the slot being formed by deposition of layers as opposed to the lithography and etching required for vertical slots. This presents a practical way of achieving slot widths down to the order of 10 nm that we have examined. Obviously this is not easily achievable with air as the slot material and is a more practical solution for the GaP-Diamond structure. If a tri-layer of GaP-diamond-GaP can be produced then it becomes easier than the air slot structures to fabricate as the smallest feature that needs to be defined by EBL, or other lithography, is the rod height $h$ which is of the order of 100 nm. In addition to this, the NV centre could be created by implantation [12] before the top GaP layer is applied.

Fabrication of the DBRs would most likely be achieved using EBL to pattern the reflectors in the same lithography step in which the slot-waveguide is defined. If it is advantageous for alignment purposes to fabricate the mirrors after the inclusion of the NV, a technique such as focused ion beam (FIB) milling may be more appropriate. A waveguide cavity using DBRs as we have discussed has been fabricated in silicon, achieving a reflectivity of 99.4% and quality factor $Q \sim 27000$ [43]. The same fabrication methods can be used to achieve alternate mirror systems such as the tapered 1D PBG structures used in [44] which have demonstrated quality factors as high as $Q = 58000$ in a waveguide cavity.

## 6. Conclusion

High optical confinement slot-waveguide cavities (SWCs) compatible with diamond optical defects (in particular, the NV) in the visible regime have been proposed in various configurations. When combined with suitable mirrors or distributed Bragg reflectors, the mode-optimized 5 nm wide GaP-air slot designs (with and without a diamond bridge) offer subwavelength-sized cavity modal volumes – nearly 2 orders of magnitude smaller than the theoretical limit of a diamond-based photonic crystal cavities, and 20-fold smaller volumes are predicted for a more modest 20 nm GaP-air slot and all-diamond designs. As a result, we expect high single-photon Rabi frequency on the order of $10^{11}$ rad s$^{-1}$ in a range of designs and slot widths. The increased light-matter interaction with diamond defect centres should relax the $Q$ requirement for achieving strong coupling conditions for enhanced photon emission and observing nonlinear effects. Given the merits of using diamond and its optically-active defects, these SWCs are enablers for quantum optical applications and new experiments in solid-state cavity-QED.

## Acknowledgments

The authors thank Photon Design support for their assistance with FIMMWAVE simulations. This project is proudly supported by the International Science Linkages programme established under the Australian Government's innovation statement Backing Australia's Ability. The authors wish to also acknowledge the Victorian Government's Science, Technology & Innovation infrastructure Grants Program for the funding of this project ADG and LCLH acknowledge the ARC for financial support (Projects No. DP0880466 and No. DP0770715, respectively).